\documentclass[galaxies,article,accept,moreauthors,pdftex]{Definitions/mdpi} 

\usepackage{graphicx}	
\usepackage{amsmath}	
\usepackage{amssymb}	
\usepackage{caption}
\usepackage{subcaption}
\usepackage{multirow}
\usepackage{blindtext}

\newcommand{\arcsec}{$^{\prime\prime}$}

\firstpage{1} 
\pubvolume{1}
\issuenum{1}
\articlenumber{0}
\pubyear{2021}
\copyrightyear{2021}
\externaleditor{Academic Editor: Francesca Loi} 
\datereceived{} 
\dateaccepted{} 
\datepublished{} 
\hreflink{https://doi.org/} 
\Title{A GMRT Narrowband vs. Wideband Analysis of the ACT$-$CL J0034.4+0225 Field Selected from the ACTPol Cluster Sample}

\TitleCitation{A GMRT Narrowband vs. Wideband Analysis of the ACT$-$CL J0034.4+0225 Field Selected from the ACTPol Cluster Sample}

\Author{Sinenhlanhla P. Sikhosana 
 $^{1,2,}$*\orcidA{}, Kenda Knowles $^{3,4}$\orcidB{}, C.H. Ishwara-Chandra $^{5}$\orcidF{}, Matt Hilton $^{1,2}$\orcidC{}, Kavilan~Moodley $^{1,2}$\orcidD{} and Neeraj Gupta $^{6}$\orcidE{}}

\AuthorNames{Sinenhlanhla P. Sikhosana, Kenda Knowles, C. H. Ishwara-Chandra, Matt Hilton, Kavilan Moodley and Neeraj Guptae}

\AuthorCitation{Sikhosana, S.P.; Knowles, K.; Ishwara-Chandra, C.H.; Hilton, M.; Moodley, K.; Guptae, N.}

\address{%
$^{1}$ \quad Astrophysics Research Centre, School of Mathematics, Statistics \& Computer Science, University of KwaZulu-Natal, Durban 4041, 
South Africa; hiltonm@ukzn.ac.za (M.H.); Moodleyk41@ukzn.ac.za (K.M.)\\
$^{2}$ \quad School of Mathematics, Statistics Computer Science, University of KwaZulu-Natal, Westville Campus,
Durban 4041, South Africa\\
$^{3}$ \quad Department of Physics and Electronics, Rhodes University, P.O. Box 94, Makhanda 6140, South Africa; kendaknowles.astro@gmail.com\\
$^{4}$ \quad South African Radio Astronomy Observatory, 2 Fir Street, Observatory, Cape Town 7405, South Africa\\
$^{5}$ \quad National Centre for Radio Astrophysics, Tata Institute of Fundamental Research, S. P. Pune University Campus, Ganeshkhind, Pune 411007, India; ishwar@ncra.tifr.res.in\\
$^{6}$ \quad Inter-University Centre for Astronomy and Astrophysics (IUCAA), Post Bag 4, Pune 411007, India; ngupta@iucaa.in}

\corres{Correspondence: SikhosanaS@ukzn.ac.za}

\abstract{Low frequency radio observations of galaxy clusters are a useful probe of the non-thermal intracluster medium (ICM), through observations of diffuse radio emission such as radio halos and relics. Current formation theories cannot fully account for some of the observed properties of this emission. In this study, we focus on the development of interferometric techniques for extracting extended, faint diffuse emissions in the presence of bright, compact sources in wide-field and broadband continuum imaging data. We aim to apply these techniques to the study of radio halos, relics and radio mini-halos using a uniformly selected and complete sample of galaxy clusters selected via the Sunyaev-Zel'dovich (SZ) effect by the Atacama Cosmology Telescope (ACT) project, and its polarimetric extension (ACTPol). We use the upgraded Giant Metrewave Radio Telescope (uGMRT) for targeted radio observations of a sample of 40 clusters. We present an overview of our sample, confirm the detection of a radio halo in ACT$-$CL J0034.4+0225, and compare the narrowband and wideband analysis results for this cluster. Due to the complexity of the ACT$-$CL J0034.4+0225 field, we use three pipelines to process the wideband data. We conclude that the experimental \textsc{spam} wideband pipeline produces the best results for this particular field. However, due to the severe artefacts in the field, further analysis is required to improve the image quality.}

\keyword{galaxies; clusters; \textcolor{yellow}{{individual--galaxies}}; clusters; \textcolor{yellow}{{intracluster medium--radio continuum}}} 

\begin{document}

\setcounter{section}{0} 

\section{Introduction}
Non-thermal diffuse emission in galaxy clusters was first discovered in the Coma cluster~\citep{1959Natur.183.1250L,2011MNRAS.412....2B,2020arXiv201108856B}. This discovery was a confirmation of the existence of relativistic electrons and magnetic fields in the intracluster medium (ICM)~\cite{1988S&T....76Q.639S,2012ARA&A..50..353K}. Traditionally, diffuse radio emissions have been categorized into three groups based on morphology, size, and cluster dynamics; giant radio halos (GRHs), radio mini-halos (RMHs), and radio relics (RRs). The size of these non-thermal diffuse structures, which extend over kpc to Mpc scales, raised questions on their formation mechanisms. One prominent question is how the cosmic ray electrons (CRes) are (re)accelerated and transported given their diffusion timescale limitations~\cite{2011A&A...527A..99E}.  Many studies have been conducted to formulate and constrain theories detailing the formation of these sources.
\par
 GRHs, Mpc-scale sources with low polarization percentages (<10$\%$) located in the central regions of clusters, have two main formation theories. The first is the secondary `hadronic' model, in which electrons originate from hadronic collisions between the long-living relativistic protons in the ICM and thermal ions~\cite{1980BAAS...12..471D,2008MNRAS.385.1211P,2011A&A...527A..99E}. This formation theory has not been widely accepted due to the lack of observational evidence of gamma rays in clusters, which are a by-product of the hadronic processes~\cite{2003ApJ...588..155R,2010PhRvD..82i2004A,2018ApJ...860...85A}. Although observations disfavour the hadronic model as the primary source of radio halo emission, hybrid models suggest that these hadronic interactions may produce a seed population of electrons that is then re-accelerated to form GRHs \citep{2005MNRAS.363.1173B,2017MNRAS.472.1506B}.

The second model is the primary `re-acceleration' model. According to this model, a pool of pre-existing electrons~\cite{2017MNRAS.465.4800P} is re-accelerated through second order Fermi mechanisms by ICM turbulence developing during cluster mergers~\cite{2001MNRAS.320..365B,2012A&ARv..20...54F,2014MNRAS.443.3564D}. A strong dynamical link has been found with respect to the host clusters~\cite{2005ApJ...627..733M,2014ApJ...786...49L,2015A&A...579A..92K}. The $\sim$Mpc-scale emission has mostly been found in massive (M$_{500,SZ}$ > 4 $\times$ $10^{14}$ M$_{\odot}$) clusters with X-ray and optical merger signatures~\cite{2013ApJ...777..141C}. The power of the radio emission has been found to correlate with thermal host cluster properties, with non-detections lying below the correlation and with sources that exhibit ultra-steep spectra ($\alpha$\endnote{S$_{\nu} \; \propto \; \nu^{-\alpha}$} $\gtrsim$ 1.5) that populate the region between the correlation and upper limits, as predicted by the re-acceleration model~\cite{2014IJMPD..2330007B}. Studies have shown that cluster selection methods affect the resulting scaling relations. Samples selected via their Sunyaev-Zel'dovich signal (SZ; \cite{1972CoASP...4..173S}) show a higher detection rate than X-ray-selected samples~\cite{2015A&A...580A..97C,2021arXiv210101641C}. This difference may be due to the different time-scales of boosting the SZ vs X-ray emission during mergers. Although the re-acceleration theory is widely accepted, there are still a few aspects that need further investigation. A major open question is the origin of the re-accelerated cosmic ray particles \citep{2014IJMPD..2330007B,2019SSRv..215...16V}.

\par
RMHs are similar in morphology to GRH; however, they extended over a few 100~kpc in scale and are generally found in non-merging cool-core clusters. These sources are located around the brightest cluster galaxies (BCGs). RMHs form as a result of re-acceleration of seed electrons by the turbulence induced from gas sloshing in the cool core~\cite{2002A&A...386..456G}. \citet{2014IJMPD..2330007B} suggest that there is an intermediate transition stage where GRHs transition to RMHs or vice-versa. This has been the explanation used for the existence of GRHs in cool-core clusters~\cite{2014MNRAS.444L..44B,2021MNRAS.504..610P}.
\par
RRs are arc-shaped, $\sim$Mpc scale, highly polarised sources ($\gtrsim$20$\%$) that are located at the peripheral region of the clusters. Several observations have shown that their origin is linked to shock waves generated in the ICM by merger events~\cite{1979ApJ...233..453J,1998A&A...332..395E,2012MNRAS.426...40B,2016MNRAS.460L..84B,2020MNRAS.tmpL..75L}. However, the underlying particle acceleration mechanism is still under debate~\cite{2019MNRAS.489.3905S}. In the mechanism of first-order diffusive shock acceleration (DSA; \cite{1978MNRAS.182..443B,1991SSRv...58..259J}), cosmic-ray protons and electrons are assumed to be accelerated from the thermal pool up to relativistic energies at the cluster merger shocks. Although this mechanism can explain the general properties of relic emission, several observational features remain unexplained~\cite{2019MNRAS.489.3905S}, such as the non-detection of gamma rays in clusters that host RRs~\cite{2018ApJ...860...85A} and the low Mach numbers observed in shocks~\cite{2014IJMPD..2330007B}. 
\par
The second mechanism proposes the re-acceleration of fossil relativistic electrons via DSA at the cluster shocks~\cite{2005ApJ...627..733M,2013MNRAS.435.1061P,2017ApJ...840...42K}. This mechanism reproduces the observed spectrum~\cite{2014MNRAS.444.3130D,2015MNRAS.449.1486S,2017NatAs...1E...5V} and does not require shocks to have large Mach numbers, as the pre-existing electrons have enough energy to be re-accelerated to relativistic speeds~\cite{2014IJMPD..2330007B}. However, this mechanism also presents challenges as there are expected phenomena that are yet to be observed~\cite{2019MNRAS.489.3905S}. For example, the connection between active galactic nuclei (AGN; candidate seed electron source) and radio relics could be established only in a few cases~\cite{2014ApJ...785....1B,2017ApJ...835..197V}. 
A model by \citet{2017NatAs...1E.163Z,2018MNRAS.478.4922Z} focuses on the role of magnetic fields that under specific configurations could allow electrons to reach relativistic speeds via the shock drift acceleration mechanism (SDA; \cite{2014ApJ...797...47G,2014ApJ...794...47C}). However, the role of magnetic fields and its amplification by low Mach number shocks is still poorly constrained~\cite{2019MNRAS.489.3905S}.
\par
A new generation of telescopes such as the LOw-Frequency ARray (LOFAR; \cite{2013A&A...556A...2V}), MeerKAT~\cite{2016mks..confE...1J}, and the uGMRT \citep{2017CSci..113..707G} have opened a window into a new group of ultra-steep radio sources which were previously undetectable due to frequency and sensitivity constraints. The new studies have resulted in a different class of diffuse emission, such as; radio phoenices and gently re-energized tails (GReETs; \cite{2017SciA....3E1634D,2019SSRv..215...16V}). The sensitivity of these telescopes have opened up a new observational window which probes lower mass \mbox{($<$4  $\times$ 10$^{14}$ M$_{\odot}$}) and higher redshift ($z$ $>$ 0.3) clusters \citep{2019MNRAS.486.1332K,2020A&A...640A.108G,2021MNRAS.500.2236R}. Large cluster samples of this nature will help in the refinement of currently existing formation theories. However, the broad bandwidths and large fields of view of these telescopes also present multiple data reduction challenges.
\par
In this paper, we introduce an SZ-selected sample from the ACTPol observations. We will also discuss the challenges we faced with the uGMRT data reduction and present a case study of ACT$-$CL J0034.4+0225 (hereafter J0034). In Section~\ref{sec-sample}, we introduce the cluster sample. In Section~\ref{sec-reduction}, we discuss the data reduction challenges and the pipelines we explored. In Section~\ref{sec-j0034}, we study J0034 and compare the narrowband and wideband results. Finally, we summarise our findings and future outlook in Section~\ref{sec-summary}. We adopt a $\Lambda$CDM flat cosmology with $H_0$ = 70 km s$^{-1}$ Mpc$^{-1}$, $\Omega_{\rm m}$ = 0.3, and $\Omega_{\Lambda}$ = 0.7. For our radio spectral index calculations, we assume S$_{\nu} \; \propto \; \nu^{-\alpha}$, where $S_\nu$ is the flux density at frequency $\nu$ and $\alpha$ is the spectral index.

\section{The ACTPol Sample}
\label{sec-sample}
Until recently, statistical studies of diffuse radio emission in galaxy clusters have been constrained to observations of high mass and low redshift systems selected using X-ray telescopes~\cite{2008A&A...484..327V,2015A&A...579A..92K}. The X-ray selected samples had relatively low diffuse emission detection rates ($\sim$20$\%$). Andrade-Santos et~al.~\cite{2017ApJ...843...76A} studied the fraction of cool-core clusters in an X-ray selected sample from Chandra 
versus an SZ selected sample from Planck. Their study revealed that the X-ray sample had a higher fraction ($\sim$44$\%$) of cool-core clusters in comparison to the SZ sample ($\sim$28$\%$). The majority of cool-core clusters do not host GRHs and RRs; and hence would attribute to the low detection rates. These findings in conjunction with the poor resolution ($\sim$1.8$'$) of ROSAT, which was mainly used for the X-ray samples, may account for the low detection rates of diffuse emission in X-ray selected samples. Such constraints have since led to preferentially using SZ-selected samples for scaling relation~studies.  

The thermal SZ effect is a powerful probe for high-redshift clusters due to the fact that it is not affected by dimming. Hence, the SZ survey catalogues offer an almost redshift independent selection function. Essentially, cluster samples from SZ surveys are restricted by cluster masses based on the sensitivity of the observing instrument. The three main microwave telescopes that have been used to detect galaxy clusters are the ACT, the Planck satellite~\cite{2011A&A...536A...1P}, and the South Pole Telescope (SPT; \cite{2011PASP..123..568C}).

\begin{figure}[H]
\widefigure
\includegraphics[width=15.0 cm]{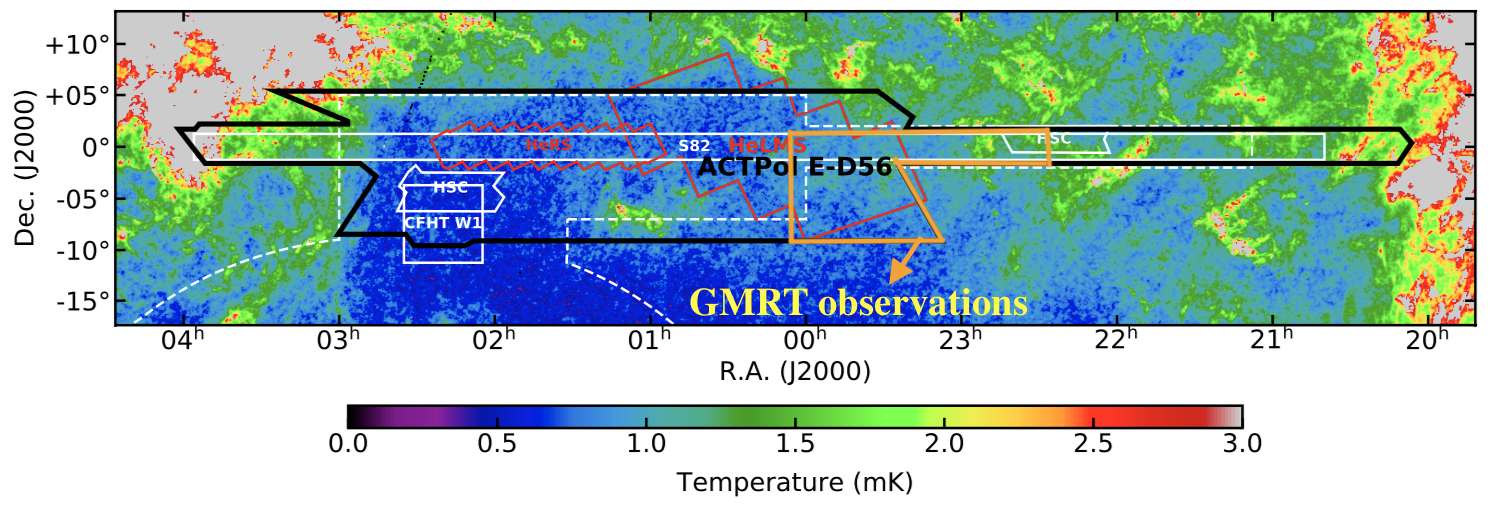}
\caption{The Planck 353 GHz temperature map overlaid with the ACTPol E$-$D56 region (shown in black) and the overlapping multi-wavelength surveys. The region enclosed by the magenta polygon indicates the region of the selected final sample, which we follow up using u/GMRT observations. Source: Adapted from Hilton et~al.~\cite{2018ApJS..235...20H}.\label{fig1}}
\end{figure}

For our project, we use the galaxy cluster catalogue from the  ACT's Polarimetric extension (ACTPol; \cite{2017JCAP...06..031L}) to select our sample. The ACTPol catalogue was constructed using the E$-$D56 region, shown in Figure~\ref{fig1}, which covers an area of 987.5 deg$^2$~\cite{2018ApJS..235...20H}. 
\end{paracol}

\begin{paracol}{2}
\switchcolumn
The final catalogue consists of 182 optically confirmed clusters. For a signal-to-noise (SNR) $>$ 4 the sample spans a mass and redshift range of 1.6 $<$ 10$^{14}$ M$_{\odot}$ $<$ 9.1, and \mbox{0.1 $<$ $z$ $<$ 1.4}, respectively. The 90$\%$ sample completeness cut-off is $M_{\rm 500c}\,>\,4.5\,\times\,10^{14}\,{\rm M}_\odot$ for SNR $>$ 5 and a redshift range of 0.2 $<$ $z$ $<$ 1.0. M$_{\rm 500c}$ is the mass measured within a radius that encloses a region with an average density that is 500 times the critical density at the cluster redshift, and assuming the SZ-signal scales with mass as described in \mbox{Arnaud et~al.~\cite{2010A&A...517A..92A}}.

For our sample, we considered all the clusters with an SNR $>$ 5. We then applied a mass and redshift cut of  M$_{\rm 500,SZ}$ $>$ 4 $\times$ 10$^{14}$\,M$_{\odot}$ and 0.1 $<$ $z$ $<$ 0.8. This results in a sample of 40 clusters, which formed the basis of our sample. Given that the ACTPol catalogue is 90$\%$ complete for clusters within 0.2 $<$ $z$ $<$ 1.0 and $M_{\rm 500c}\,>\,4.5\,\times\,10^{14}\,{\rm M}_\odot$, we derive our sample completeness to be 68$\%$. This completeness is derived from the mass completeness and the available radio information (30/40). In Figure~\ref{fig2a}, we compare our sample of 40~clusters to \textit{Planck} and SPT samples, which are the most recent statistical studies using SZ-selected cluster samples. Our sample covers a wider redshift range compared to both samples. We also cover lower mass clusters compared to the \textit{Planck} sample. The SPT sample overlaps with ours at higher redshifts; however, it does not cover lower redshifts \mbox{($z$ $<$ 0.33)}. This study is a precursor to studies of larger samples exploring lower mass and higher redshift cluster sample studies such as the MeerKAT Extended Relics, Giant Halos, and Extragalactic Radio Sources (MERGHERS) survey~\cite{2016mks..confE..30K} and the MeerKAT Absorption Line Survey (MALS; \cite{2016mks..confE..14G}).

We chose to observe these clusters at low-frequency (250--500 MHz) using the wideband uGMRT in addition to using existing archival narrowband GMRT data. As of November 2020, there were 17 clusters with pre-existing GMRT legacy software backend (GSB) observations. The GSB has a bandwidth of 32 MHz. We obtained uGMRT wideband (GWB) observations for 13 clusters, taken over three observing semesters. These clusters were observed in GMRT's observing cycles 32, 33, and 36 (proposal IDs: 32$\_$012, 33$\_$010, 36$\_$050). In total, we have 30 clusters with narrowband and/or wideband observations. For this paper, we focus on the data reduction and analysis of J0034, which has both narrowband and wideband observations. We selected this field because it was most severely affected by RFI and has multiple bright sources near the cluster region. Hence, successfully calibrating and imaging this field would imply that we could apply the techniques on the rest of the~sample. 

\begin{figure}[H]
\includegraphics[width=9.6 cm]{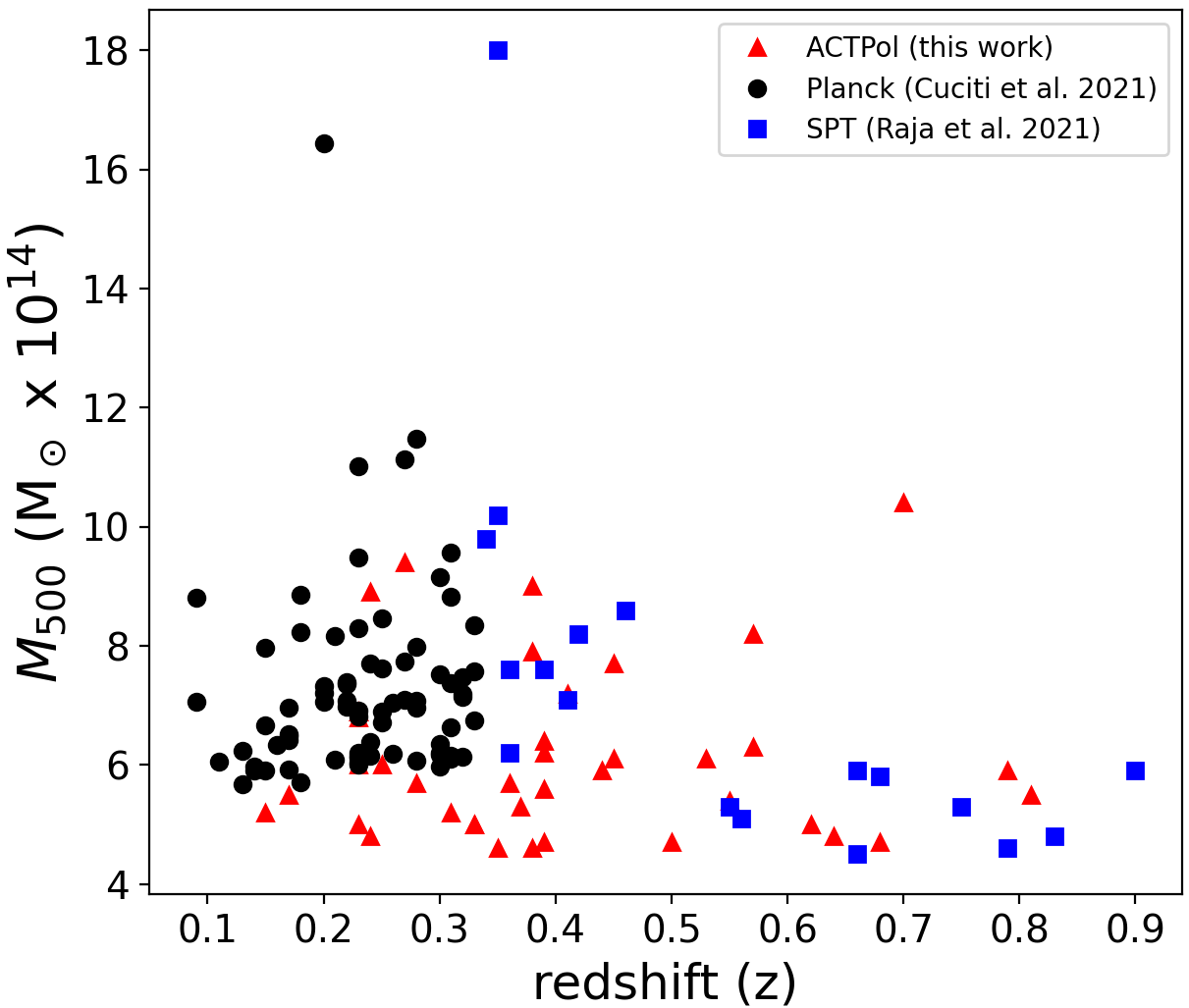}
\caption{The ACTPol sample in comparison to the Planck sample~\cite{2021arXiv210101641C} and SPT~\cite{2021MNRAS.500.2236R} sample used for statistical studies of diffuse emission.\label{fig2a}}
\end{figure}  
\section{Data Reduction}
\label{sec-reduction}
For this paper, we focus on the case study of the complex field J0034. We start by reducing the narrowband observations using the source peeling and atmospheric modelling (SPAM; \cite{2014ASInC..13..469I,2017A&A...598A..78I}). The final narrowband image indicates that this field has bright sources near the cluster's region. Such sources complicate the calibration procedure. Such a field is a good case study for testing calibration and imaging strategies. Hence, when reducing the wideband data, we use three pipelines and compare the resulting images for J0034. In the following sections, we discuss the narrowband and wideband data reduction strategies.
\subsection{Legacy GMRT GSB}
 We use \textsc{spam}'s calibration pipeline version 19.11.07 to reduce the GSB data. \textsc{spam} is a fully automated Python pipeline that uses the \textsc{parseltongue}~\cite{2006ASPC..351..497K} interface to access and execute the Astronomical Image Processing System (AIPS; \cite{1996ASPC..101...37V}) tasks. The pipeline is divided into two steps; the `pre-calibrate targets' step and the `process target' step. The data reduction steps are as follows.
\begin{enumerate}
\item The `pre-calibrate target' step performs the cross-calibration step of the standard calibration procedure. The pipeline uses the primary calibrator(s) to determine the channels affected by RFI, determine the flux scale, and to produce cross calibration tables. The flags and calibration tables are then applied to the target source. Finally, the calibrator and target visibilities are split into separate UVFITS files.  

\item The `process target' step begins by taking the cross-calibrated (1GC) target data and applying 2GC (or self calibration). For 2GC, the target visibilities are imaged using the facet-based method established by \citet{1992A&A...261..353C}. The point sources covering the primary beam are obtained from a sky model extracted from the VLA low-frequency sky survey (VLSS; \citep{2007AJ....134.1245C}) and the NRAO VLA sky survey (NVSS; \citep{1998AJ....115.1693C}). The observed field is then faceted based on the sky model. The deconvolution is done using the Cotton-Schwab \textsc{clean} algorithm \citep{1998AJ....115.1693C}. The \textsc{cleaned} visibilities are calibrated using the sky model. Then the calibrated visibilities are imaged to produce a better sky model and improved calibration solutions. This self calibration cycle is applied three times. For the wide-field imaging, the pipeline performs a single-scale CLEAN deconvolution
down to 3 times the central background noise ($\sigma$), using automated CLEAN boxes placed at positive peaks of at least 5$\sigma$.

\item The second part of the `process target' step  is to correct for DDEs (3GC), which include ionospheric effects \citep{2005ASPC..345..399L}. For each 3GC cycle, the brightest source in the field is peeled. The ionospheric effects of the brightest source are modelled and phase corrected solutions for the peeled model are obtained. \textsc{spam} uses a phase screen model to extend the ionospheric phase solutions of the single source to the full field. Then the solutions are applied to each facet during imaging. This cycle is repeated six times, and each time the brightest source is selected and corrected for DDEs. This repeats until all the bright sources are corrected for DDEs.  Finally, the astrometric corrections are applied to the DDEs corrected image and the image is primary beam corrected thereafter.

\end{enumerate}

We used default data reduction parameters throughout the different stages of the pipeline. The default setting of the number of pixels is $\sim$3780, with a pixel size of $\sim$1.9$''$ and the robust parameter is~1. The resulting image, shown in the bottom right panel of Figure~\ref{fig:j0034-pipe}, has a rms of 0.10 mJy/beam near the cluster centre, and the beam size is \mbox{18.1\arcsec $\times$ 14.2\arcsec,} p.a. 60.3$^{\circ}$.

\begin{figure}[H]
\includegraphics[width=13.5cm]{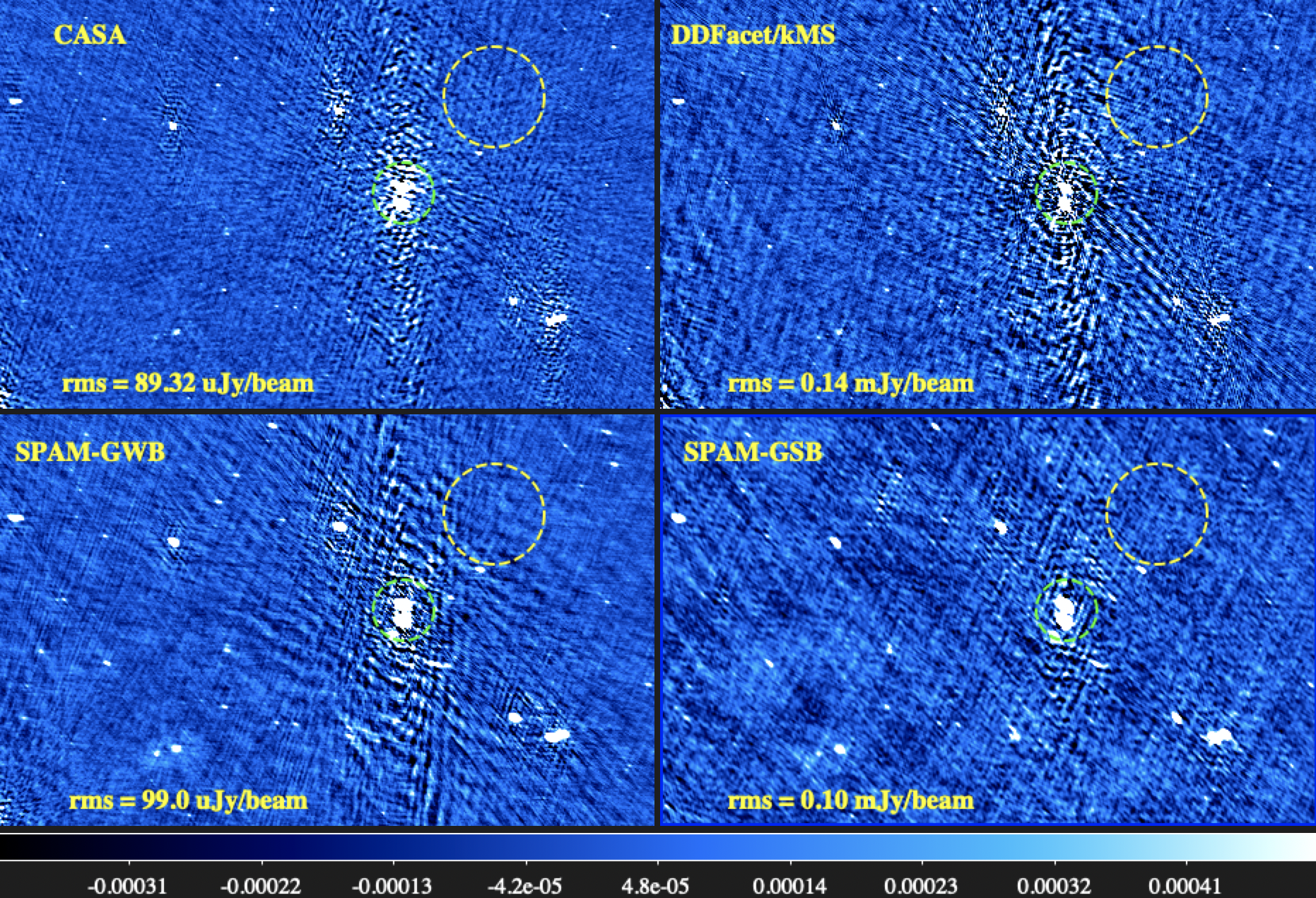}
\caption{Gallery of GWB pipeline images of the J0034 field, which indicates the severity of the DDEs in our observations. The artefacts in all the frames are detected at 2$\sigma$. The lower left panel shows the GSB image for comparison. In each panel, the green circle indicates the source used to measure peak brightness, and the yellow circle indicates the region, with the rms value indicated at the lower left. The angular radius of the noise region is 1.5$'$. The colour scale is the same for all panels. 
\textbf{Top left:} 
The CASA-pipeline reduced image (see Section~\ref{casa-pipe}). The beam size of the image is the beam size is (8.0\arcsec $\times$ 4.9\arcsec, p.a. 60.3$^{\circ}$). \textbf{Top right:} The DDFacet$-$killMS$-$pipeline reduced image (see Section~\ref{sec-ddf}). The beam size of the image is (7.9\arcsec $\times$ 4.5\arcsec, p.a. 52.4$^{\circ}$) \textbf{Bottom left:} Experimental wideband \textsc{spam}-pipeline reduced image (see Section~\ref{sec-expspam}). The beam size of the image is (\mbox{14.0\arcsec $\times$ 7.6\arcsec}, p.a. 65.3$^{\circ}$). \textbf{Bottom right:} The 325 MHz narrowband image produced using the standard \textsc{spam} pipeline. The beam size of the image is (18.1\arcsec $\times$ 14.2\arcsec, p.a. 48.2$^{\circ}$).}
\label{fig:j0034-pipe}
\end{figure}

\subsection{uGMRT Data Reduction}
\textls[-25]{Thirteen clusters in our sample, including J0034, have uGMRT GWB band 3 (\mbox{250--500 MHz})} observations. The wide bandwidth of the upgraded GMRT simultaneously provides increased sensitivity and opportunity for in-band spectral index studies. However, the calibration and imaging of GWB data is a complex procedure. The main challenges with reducing the wideband, wide-field data are as follows. The primary beam pattern is dependent on the observing frequency, hence, for wide bandwidth observations, this pattern varies across the band. The flux density of radio sources correlates with the frequency (S$_{\nu}\,\propto\,\nu^{-\alpha}$). As a result of this correlation, the spectral models of the point sources need to be taken into account during imaging. The uGMRT observations were carried out at low-frequencies, which resulted in various directional dependent effects, such as ionospheric effects. The GWB observations of J0034 are particularly affected by radio frequency interference (RFI), which corrupts data if not fully modelled and removed. The level of RFI in our data makes the reduction process onerous and leads to images with significantly compromised quality. The resulting images had rms noise values three to seven times higher than the theoretically predicted noise floor of 10 $\upmu$Jy/beam. Another challenge for our sample was that it is in the equatorial region ($-$7.2$^{\circ}$ $<$ $\delta$ $<$ 4$^{\circ}$), so despite the wide bandwidth and hours of integration time, $uv$ coverage is more sparse because aperture synthesis is not as effective for equatorial sources (see Figure~\ref{fig:j0034-uv}). The poor sampling of the visibility space results in north-south artefacts around bright sources, which are a reflection of the poor sampling function. 
\begin{figure}[H]
\includegraphics[width=12.5 cm]{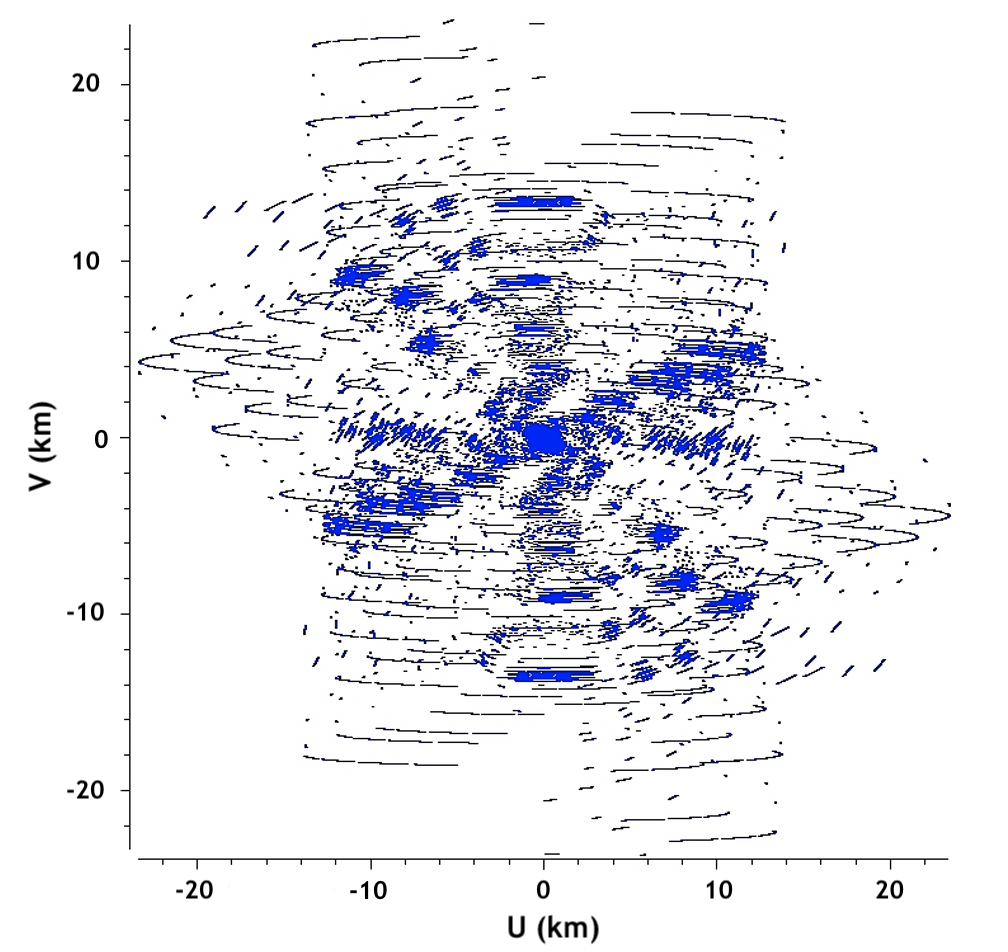}
\caption{GMRT \emph{uv} coverage for J0034. Despite the 4 h on-source and the 200 MHz of bandwidth, the equatorial declination results in sparse \emph{uv}$-$coverage which reduces the effectiveness of aperture synthesis.}
\label{fig:j0034-uv} 
\end{figure}

\par
To overcome these challenges, we explored various pipelines that deal with the intensive RFI flagging and calibration direction-dependent effects. Once the data reduction pipelines are refined, we plan to reduce the remaining wideband data to determine if we can detect lower surface brightness diffuse emission in clusters with non-detections in the GSB data, and enhanced features of the existing diffuse emission detections in other clusters. The pipelines we explored are summarised below, including the image comparison of the J0034 results.
\subsubsection{CASA Pipeline}
\label{casa-pipe}
The \textsc{casa} pipeline we adopt is a preliminary version of the \textsc{capture} pipeline~\cite{2020ascl.soft11002K}. For this pipeline, flagging, calibration and imaging is done using in-built Common Astronomy Software Applications (\textsc{casa}; \cite{2007ASPC..376..127M}) tasks. Firstly, we flag RFI using the manual and automated \texttt{flagdata} tasks. We apply the auto-flag separately for the different fields. For auto-flagging, we use the \texttt{tfcrop} mode which identifies and removes outliers on the time-frequency plane. We set the flagging threshold parameters higher for the calibrator fields in comparison to the target field because these fields are usually much brighter and are detected at a higher SNR. The \texttt{timecutoff} and \texttt{freqcutoff} deviation parameters for the calibrators were set to 5.0, while for the target they were set to 6.0. These values place constraints on the deviation of data points from the fitted time and frequency polynomial. Data points with higher deviation values are regarded as RFI. 
\par
We then apply a cycle of cross-calibration. We begin by setting the flux scale using the standard models from~\cite{2013ApJS..204...19P}. Thereafter, we produce calibration tables. The delay calibration solution interval is 10 min, and we only have one solution interval for bandpass and gain calibration. We transfer the flux and phase calibration solutions to the target. After the first cycle of cross-calibration, we apply flagging once more. We do this in order to excise low-level RFI only discernible after calibration. This time we use \texttt{tfcrop} and \texttt{rflag} mode. The \texttt{rflag} mode calculates statistics per time chunk and set thresholds which indicate the outliers that need to be flagged. We apply \texttt{rflag} post calibration because it tends to result in higher flag percentages if applied on uncalibrated data. Finally, we produce calibration solutions from the second cycle of cross-calibration and apply these solutions to the target.
\par
We then separate the target visibilities and begin with 2GC. We apply multiple cycles of phase-only calibration on the data, we stop once the quality of the image is no longer improving. For most data sets, the four cycles were sufficient. We then apply multiple cycles of phase and amplitude calibration until the noise quality of the images reaches a plateau, three cycles were sufficient for most datasets. For both phase-only and phase and amplitude calibration, we begin with a solution interval of 16 min and decrease the interval per cycle by dividing the initial solution interval by a factor of two times the number of the cycle. Our imaging during self-calibration is done using \texttt{tclean}'s \texttt{mtmfs} deconvolver, with nterms = 2 and robust parameter = 0. This deconvolver accounts for the varying spectral indices of sources across the wide bandwidth. The number of iterations is set to 2500 and increases per cycle by a factor of 2$^n$, where n is the cycle number. The mask is set to auto-multithreshold. The CLEANing threshold is 0.01 mJy while the sidelobe threshold is set to 3$\sigma$. 
\par
The resulting image, shown in the top left panel of Figure~\ref{fig:j0034-pipe}, has a rms of 89.2 $\upmu$J/beam and the beam size is 8.0\arcsec $\times$ 4.9\arcsec, p.a. 60.3$^{\circ}$. The GSB image is included in the panels for comparison purposes. After rigorous RFI excision strategies, including \textsc{aoflagger}~\cite{2012A&A...539A..95O}, the north--south artefacts around the bright sources were still visible. This led us to conclude that these artefacts and phase variations in the bright sources are due to DDEs. The results of this pipeline were unsatisfactory for our science goals, since we needed to correctly remove bright point sources in the visibilities. Hence, we explored other pipelines which apply DDEs calibration (3GC).

\subsubsection{DDFacet/killMS Pipeline}
\label{sec-ddf}
The \textsc{killms}\endnote{\url{https://github.com/saopicc/killMS}} (the pipeline was accessed in August 2019) and 
\textsc{ddfacet} \citep{2018A&A...611A..87T} based pipeline attempts to correct for DDEs by solving the full Jones matrix. This pipeline starts with a measurement set (MS) that has been calibrated using the \textsc{casa} pipeline. From our investigation of the \textsc{casa} pipeline, we found that more cycles of 2GC calibration resulted in images with lower peak fluxes. Hence, for this pipeline, 2GC is only applied once for phase-only calibration. We use the image produced after 2GC to create a mask using a threshold of 10$\sigma$ to ensure we create a sky model that does not contain residual artefacts. Thereafter, we image the visibilities using \textsc{ddfacet}. The imaging parameters used are as follows. We apply three major cycles and 90,000~minor cycles with a deconvolution peak factor of 0.001. These parameters are fixed even for the post-\textsc{killms} imaging cycle.
\par
We then examine the 2GC image and locate the brightest sources ($>$0.1 Jy) in \textsc{ds9}~\cite{2003ASPC..295..489J}, and create a bright-source region file. A minimum of three sources is tagged for each field, and a maximum of 6 sources are tagged for fields with numerous bright sources. We use the region file to divide the target field into facets equal to the number of bright sources in the region file, with the tagged sources being at the phase centre of each facet. We use \textsc{killms} to obtain a set of direction-independent solutions for each facet separately, which is combined into a set of direct-dependent solutions for the field. We apply the \textsc{CohJones} solver with a time solution interval of 5 min and frequency interval of 8 channels per solution. The number of directions we solve for is equivalent to the number of the bright sources that are tagged per field. Finally, we use \textsc{ddfacet} to apply the \textsc{killms} solutions and produce the DDEs corrected image. \textsc{ddfacet} also applies direction-dependent PSF deconvolution, which explicitly accounts for the time variation and bandwidth fluctuation effects. We perform the \textsc{killms} and \textsc{ddfacet} loops iteratively, improving the mask and increasing the time solution intervals with each loop until we get an image with significantly reduced artefacts. 
\par
The resulting image, shown in the top right panel of Figure~\ref{fig:j0034-pipe}, has a rms of \linebreak0.14~mJy/beam and the beam size is 7.9\arcsec $\times$ 4.5\arcsec, p.a. 52.4$^{\circ}$. We note that the noise floor is higher than the\textsc{casa} result. The noise for the J0034 field increases by $\sim$4$\%$. However, the sources in this pipeline have a better defined structure compared to \textsc{casa}. The point sources are circular and have less phase variation. This indicated that the pipeline improved the phase corrections for the fields. Although the phase corrections had improved in comparison to the \textsc{casa} pipeline, the north--south artefacts were still not significantly reduced. Such artefacts would have been problematic for the point source subtraction, which is often required when extracting measurements for faint extended emission. We tried various calibration solution intervals (30 s--2 min)  and facet numbers (3--8); however, these did not improve our results. The range of calibration solution intervals produced the same results, while increasing the facet numbers resulted in some facets having higher noise levels due to fewer sources in each facet.

\subsubsection{Experimental SPAM Wideband Pipeline}
\label{sec-expspam}
The final pipeline we explored was the experimental wideband \textsc{spam}\endnote{\url{http://www.intema.nl/doku.php?id=huibintemaspampipeline}} (the pipeline was accessed in October 2020). This pipeline begins by splitting the GWB data into $\sim$7 sub-bands of $\sim$30 MHz each. Thereafter, it follows the conventional GSB narrowband data reduction for the individual sub-bands (see Section~\ref{sec-reduction}). The calibrated sub-band visibilities are then converted into MS files and concurrently imaged using \textsc{wsclean}. For the \textsc{wsclean} wide bandwidth imaging step, we use a Briggs robust = 0, number of iterations = 150,000, threshold = 1 $\upmu$Jy, \mbox{auto-mask = 9$\sigma$}, amd multiscale scales of (10,20,30).

\textls[-15]{The resulting image, shown in the bottom left panel of Figure~\ref{fig:j0034-pipe}, has a rms of \mbox{99.0 $\upmu$J/beam}} and the beam size is 14.0\arcsec $\times$ 7.6\arcsec, p.a. 65.3$^{\circ}$. As seen in the bottom left panel of \mbox{Figure~\ref{fig:j0034-pipe}}, this pipeline produces the best phase calibration, which results in an improvement in the structure of the brightest sources. The flag percentages in Table \ref{tabj0034-pipe} also show that the \textsc{spam} pipeline results in the least flag percentages. These low percentages result in higher sensitivity, and this is indicated by the higher peak flux recovered for the double-lobed source shown in Figure~\ref{fig:j0034-pipe}.

\end{paracol}
\nointerlineskip
\begin{specialtable}[H]
    \tablesize{\small}
    \widetable
\caption{J0034 GWB data reduction pipelines' image comparison. Columns: (1) Name of pipeline. (2) Effective observing frequency. (3) Synthesised beam of the image. (4) Flagged data. (5) Overall rms of the full resolution images. (6) Peak brightness of the double lobed source.} \label{tabj0034-pipe}
\setlength{\cellWidtha}{\columnwidth/6-2\tabcolsep+0.0in}
\setlength{\cellWidthb}{\columnwidth/6-2\tabcolsep+0.0in}
\setlength{\cellWidthc}{\columnwidth/6-2\tabcolsep+0.0in}
\setlength{\cellWidthd}{\columnwidth/6-2\tabcolsep+0.0in}
\setlength{\cellWidthe}{\columnwidth/6-2\tabcolsep+0.0in}
\setlength{\cellWidthf}{\columnwidth/6-2\tabcolsep+0.0in}
\scalebox{1}[1]{\begin{tabularx}{\columnwidth}{>{\PreserveBackslash\raggedright\arraybackslash}m{\cellWidtha}>{\PreserveBackslash\centering}m{\cellWidthb}>{\PreserveBackslash\centering}m{\cellWidthc}>{\PreserveBackslash\centering}m{\cellWidthd}>{\PreserveBackslash\centering}m{\cellWidthe}>{\PreserveBackslash\centering}m{\cellWidthf}}
\toprule 
\multirow{2}{*}{\bf Pipeline} & \boldmath$\nu_o$ & \textbf{Flags} & \textbf{Synthesised Beam}  &\textbf{RMS} & \textbf{Peak Brightness}  \\
 & \textbf{MHz}   & \boldmath$\%$ & \textbf{\boldmath\arcsec \boldmath$\times$ \arcsec, PA(\boldmath$\textdegree$)}&\boldmath$\upmu$\textbf{Jy/beam} & \textbf{Jy/beam} \\ \midrule
\textsc{casa} &397& 46.8 & 8.0 $\times$ 4.9, 60.3  & 38.9 &0.13 \\
\textsc{ddfacet}/\textsc{killms}  & 397 &35.6  & 7.9 $\times$ 4.5, 52.4& 40.5 & 0.19 \\
\textsc{spam} GWB & 398 &  24.4 & 14.0 $\times$ 7.6, 65.3 & 45.9& 0.27 \\
\textsc{spam} GSB & 323 & 14.7 &18.1 $\times$ 14.2, 48.2 & 75.8& 0.34 \\
\bottomrule    
\end{tabularx}}
\end{specialtable}
\begin{paracol}{2}
\switchcolumn

Although the noise levels are higher compared to the \textsc{casa} pipeline for both images, the improvement of the phase calibrations meant we would be able to extract the bright point sources, which is a necessary step when extracting faint diffuse emission. The peak fluxes of the sources in the \textsc{spam}-reduced images were much higher in than those in the \textsc{casa}-reduced and \textsc{ddfacet}/\textsc{killms}-reduced images. This indicates that the phase and amplitude calibration is improved, despite the higher noise floor, so that less of the real signal is being fractured into artefacts or side lobes. However, further steps are needed to improve the algorithm so that the 3GC strategy is more robust. These steps include more comprehensive ionospheric corrections in the sub-band data reduction and primary beam correction for the full bandwidth image.

\subsection{Flux Density and P$_{\rm 1.4GHz}$ Calculations}
\label{sec:flux}
We use the 2GC calibrated  data to search for extended diffuse radio emission in the cluster region. To enhance the brightness of the diffuse emission, we produce a low resolution image. For the low resolution imaging process, we use the \textsc{spam} pipeline. We first image the compact sources by applying a $uv$-range cut at baselines $\geq$ 3 k$\lambda$. At this $uv$-range, all the point source emission is captured, while higher $uv$ cuts result in residual point source emission and lower $uv$ cuts also capture the extended diffuse structure. We then use the high resolution imaging step to create model visibilities that only contain point sources. For the high resolution image, we use a Briggs robust parameter of $-$0.8 and a $uv$-range $\gtrsim$ 6 k$\lambda$. We subtract the point sources in the visibility plane. We then produce a full resolution point-source-subtracted image to ensure that the subtraction is successful. Finally, we image the visibilities in the $uv$-range $\leq$ 8 k$\lambda$ whilst applying a $uv$-taper at 5~k$\lambda$ and using a Briggs robust parameter of 0.8. These tapering and $uv$-cut values are ideal for capturing the full extension of the emission while producing an image with decent resolution and rms noise levels for the J0034 field. 
\par
We followed the same procedure for the GWB dataset. The point sources are subtracted in the sub-band datasets, and the point-source-subtracted MS files are imaged concurrently using \textsc{wsclean}. Finally, we proceed to extract the flux density of the detected diffuse radio source. We use the statistical method derived in Knowles et~al.~\cite{2016MNRAS.459.4240K} to account for the contamination from the point source subtraction procedure. We select~100 random off-source positions and create a source catalogue using the high-resolution image. We use the off-source and source positions to calculate flux densities on beam-sized regions in the low resolution image. We calculate the mean ($\mu$) and standard deviation ($\sigma$) of the flux densities. The flux contamination resulting from the point source subtraction is the mean of the on-source positions ($\mu_{srcs}$). This results in a flux measurement given by
\begin{equation}
S = S_{meas} - (\mu_{srs} \times N) \;,
\end{equation}
where $S_{meas}$ is the measured flux density and $N$ is the number of beams within the region that the flux density is measured.
We use a polygon region guided by 3$\sigma$ contours to measure the flux density of the detected diffuse emission in the low resolution image. The systematic error due to point source subtraction is 
\begin{equation}
\sigma_{syst}^2 = \sigma_{srcs}^2 - \delta_{off-src}^2 \;,
\end{equation}
\textls[-15]{where $\sigma_{srcs}$ and $\sigma_{off-src}$ are the standard deviations of the on-source and off-source populations, respectively. We determined the uncertainty associated with the flux measurement~as}
\begin{equation}
\Delta S = \sqrt{(\delta S \times S)^2 + N (\sigma_{rms}^2+\sigma_{syst}^2}) \;,
\end{equation}
where $\delta S$ is the calibration uncertainty ($\sim$10$\%$ for GMRT \citep{2004ApJ...612..974C}), $\sigma_{rms}$ is the rms noise of the low resolution images, measured using a background region tagged in \textsc{ds9}, $\sigma_{syst}$ is the systematic error from the point source subtraction, and $N$ is the number of beams within the region of flux measurement. 
\par
Since J0034 has archival multi-frequency radio observations, we compute the integrated spectral index value using $S_{\nu} \propto \nu^{-\alpha}$, where $\nu$ is the central frequency of the observation and $S_{\nu}$ is the measured flux at that frequency. The error associated with the spectral index is calculated arithmetically, with the assumption that the flux-density measurement errors at the two frequencies are independent. We use the spectral index value to extrapolate the radio power at 1.4 GHz. We use the following equation to calculate the k-corrected radio power at 1.4 GHz
\begin{equation}
\frac{P_{1.4 \rm GHz}}{\rm WHz^{-1}} = 4\pi  \left(\frac{D_L}{\rm m}\right)^2  \left(\frac{S_{\nu}}{\rm m^{-2} \rm WHz^{-1}}\right) \left(\frac{1.4\, \rm GHz}{\nu}\right)^{-\alpha} (1+z)^{-(-\alpha+1)} \;,
\end{equation}
where $D_L$ is the luminosity distance of a cluster at redshift $z$, $S_{\nu}$ is the flux density at frequency $\nu$, and $\alpha$ is the power law spectral index. We also measure the largest angular size (LAS) and the largest linear size (LLS) of the diffuse emission. The LLS is calculated by dividing the LAS by the angular size distance ($D_{A}$). In the following section, we present a case study of J0034 which was reported to host diffuse radio emission by Knowles~et~al.~\cite{2020arXiv201215088K} (hereafter K20), and we compare the GSB and the GWB results for this cluster. 

\section{Case Study: ACT-CL J0034.4+0225}
\label{sec-j0034}
J0034 is at redshift 0.382 and has a mass of $M_{\rm 500c}$ = 9.0 $\times$ 10$^{14}$\,M$_{\odot}$~\cite{2018ApJS..235...20H}. At this cluster redshift, 1$^{\prime\prime}$ corresponds to 5.223 kpc and the luminosity distance is 2057.4 Mpc. \mbox{Carrasco et~al.~\cite{2017ApJ...834..210C}} used data from the FOcal Reducer and low dispersion Spectrograph 2 \mbox{(FORS2; \citep{1998Msngr..94....1A})}, which is mounted at Very Large Telescope (VLT) to study the galaxy cluster. From their observations they derived that the cluster has a rest-frame velocity dispersion of 713 $\pm$ 179 kms$^{-1}$ and a dynamical mass of (3.37 $\pm$ 2.19)\,$\times$ 10$^{14}$\,M$_{\odot}$. MeerKAT L-band observations by K20 indicated that the cluster hosts a candidate radio halo. From their observations, they measured the flux density of the radio halo to be \mbox{$S_{\rm 1.16GHz} =$ 1.26 $\pm$ 0.2 mJy}, and it has LLS  of 348 kpc.
\par
From our 325 MHz GSB data (PI: Kenda Knowles, ID: 32$\_$016), we observe the diffuse emission for the first time at this frequency, and confirm its presence as tentatively detected at 1.16\,GHz by K20 (see Figure~\ref{j0034-compare}). The emission we detect has a larger extent than that detected in K20 due to the observations being at a lower frequency. The flux density of the detected diffuse emission is $S_{\rm 323MHz} =$ 16.77 $\pm$ 2.34 mJy, and the LAS is 2.32$'$ $\times$ 1.46$'$, corresponding to a LLS of 726 kpc $\times$ 459 kpc. We note that the flux density measurements might contain residual emission from the BCG that is embedded in the radio halo. This contamination is accounted for as explained in Section~\ref{sec:flux}. Using our results and the 1.16\,GHz flux density measurement from K20, we calculated the spectral index of the radio halo and found it to be 2.26 $\pm$ 0.32. This indicates that the radio halo is an ultra steep spectrum. We used the spectral index to extrapolate the halo radio power at 1.4 GHz and found it to be (0.60 $\pm$ 0.15) $\times$ 10$^{24}$ WHz$^{-1}$. Given the extent of the diffuse emission, the central location, and its regular morphology, we confirm that it falls in the category of radio halos. We also note that the irregular morphology of the emission might be an indication of merger activity, given that this cluster is paired with ACT$-$CL J0034.9+0233 which is at the same redshift with a projected separation of 3.5 Mpc~\cite{2018ApJS..235...20H}.

From the GWB data (PI: Kenda Knowles, ID: 32$\_$016), we detect extended diffuse emission in the central region of the cluster. We measure the LAS of the diffuse emission to be 2.31$'$ $\times$ 1.48$'$, corresponding to a LLS of 723 kpc $\times$ 463 kpc. This is similar is size to the GSB detection (LLS $\sim$ 726 $\times$ 459 kpc). The sizes and morphology of the diffuse emission are similar for both the GWB and the GSB images  (See Figure~\ref{j0034-compare}). Hence, we retain the radio halo classification from the narrowband data analysis. 

\textls[-15]{The flux density of the radio halo for the GWB observations is \mbox{$S_{\rm 398MHz} =$ 6.22 $\pm$ 1.64 mJy}.} We use this value and the 1.16\,GHz flux density measurement from K20 to calculate the spectral index of the radio halo. We determine a spectral index of $\alpha^{1160}_{398} = 1.75 \pm 0.36$, consistent within the uncertainties of that determined using the GSB measurement. Using our GWB flux density and measured spectral index, we determine a 1.4 GHz radio power of (0.59 $\pm$ 0.12) $\times$ 10$^{24}$ WHz$^{-1}$, in agreement with the GSB extrapolated value. We recover the radio halo at a higher SNR in the GWB data. The halo radio power is in agreement for both the GWB and GSB derived values. The morphology of the radio halo in the narrowband image is slightly elongated compared to the wideband image. The difference in morphology could be an indication that the emission at the edge of the halo is faint and hence picked up by the GWB observations, which goes down to frequencies of $\sim$250 MHz.
\newpage

\end{paracol}
\begin{figure}[H]
\widefigure
\includegraphics[width=8.4 cm]{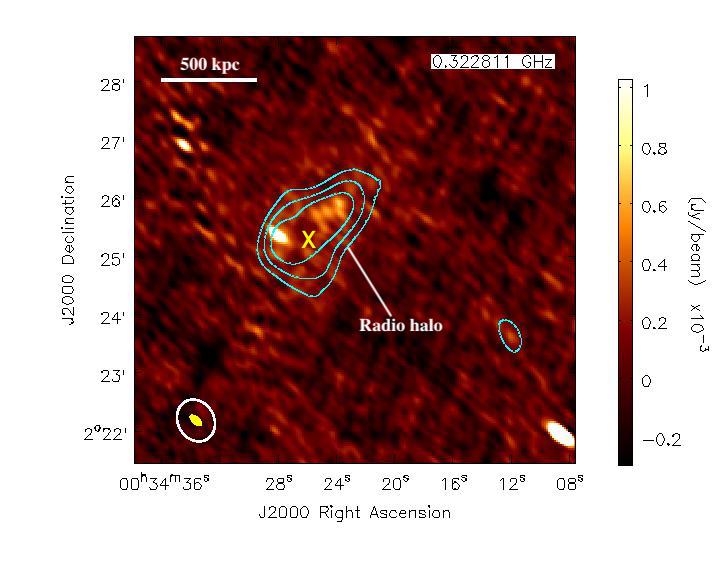}
\includegraphics[width=9.8 cm]{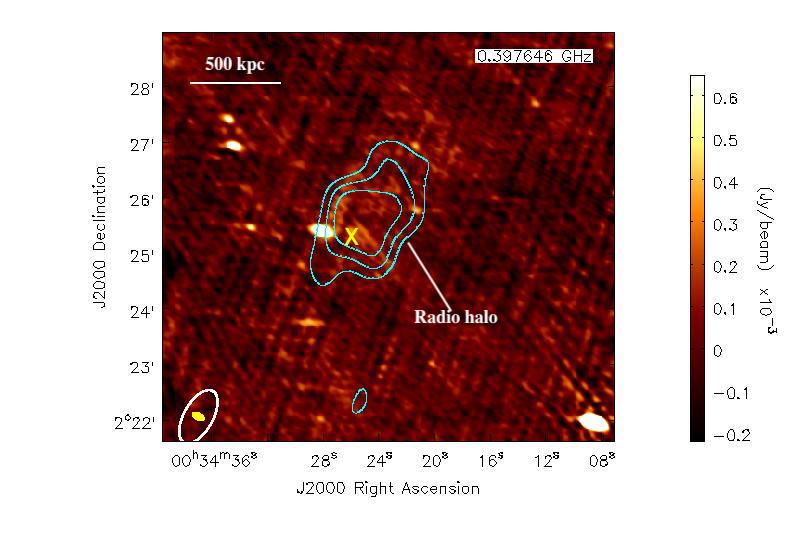}
\vspace{-6pt}
\caption{\textbf{Left:} 
Full resolution GSB image of J0034, which has a rms of 75.8 $\upmu$Jy/beam. The contours are from the low resolution point-source-subtracted image. The rms ($\sigma$) of the low resolution image is 0.58 mJy/beam and the contour levels are $\sigma$ $\times$ [$-$3,3,4,6]. The yellow cross indicates the ACT SZ peak. The beam sizes of the full resolution and the low resolution images are shown at lower left by the yellow and white ellipses, respectively. The beam sizes of the full resolution and low resolution images are (18.1\arcsec $\times$ 14.2\arcsec, p.a. 48.2$^{\circ}$) and (44.6\arcsec $\times$ 37.1\arcsec, p.a. 28.6$^{\circ}$), respectively. \textbf{Right:} Full resolution GWB image of J0034 from the experimental \textsc{spam} pipeline. The rms in the cluster region is 45.9 $\upmu$Jy/beam. The contours are from the low resolution point-source-subtracted image. The rms of the low resolution image is 0.29 mJy/beam and the contour levels are $\sigma$ $\times$ [$-$3,3,5,8]. The yellow cross indicates the ACT SZ peak. The synthesised beam for the full resolution and the low resolution images are shown by the yellow ellipse and the white ellipse, respectively. The synthesised beam for the full resolution and low resolution images are (14.0\arcsec $\times$ 7.6\arcsec, p.a. 65.3$^{\circ}$) and (43.5\arcsec $\times$ 31.1\arcsec, p.a. 150.7$^{\circ}$), respectively. \label{j0034-compare}}
\end{figure} 
\begin{paracol}{2}
\switchcolumn

For this cluster, the GWB data did not significantly add any new scientific information for the radio halo. However, we note that this dataset was severely affected by RFI and DDEs. The bright sources make this field an extreme data reduction challenge and is therefore a good candidate to test future SKA pipelines.

\section{Summary and Future Outlook}
\label{sec-summary}
The sensitivity of new generation telescopes has allowed for the study of diffuse emission in previously unexplored parameter spaces. Recent statistical studies of diffuse radio emission now consist of cluster samples that target lower mass and higher redshift clusters. Such studies are crucial for understanding the cosmological evolution of the diffuse radio sources, their connection to the dynamical state of the host cluster, and for the refinement of the currently existing formation theories.
\par
In this paper, we presented an overview of an SZ-selected ACTPol sample of 40~clusters that spans a wide mass (4.5 $<$ 10$^{14}$ M$_{\odot}$ $<$ 10.5) and redshift range (0.15 $<$ $z$ $<$ 1.0). We followed up this sample using the GMRT. Seventeen clusters have existing archival GSB observation, and we obtained uGMRT GWB band 3 data for thirteen clusters. We reduced the J0034 GSB data using \textsc{spam}. We then presented the challenges we faced with reducing the GWB data for the complex J0034 field, and the three pipelines we explored. The experimental \textsc{spam} wideband pipeline produced the most improved results. Finally, we used J0034 as a case study to compare the GSB and GWB data analysis results. We found that halo radio power is in agreement for both the GWB and GSB derived values. Although the GWB data did not significantly add any new scientific features for the radio halo, the flux is detected at a higher $\sigma$ level. We note that the uGMRT data reduction still needs to be refined to perform in-band spectral index studies.
\par
We note that the current best pipeline for the GWB data is still in the experimental stage and could be optimised to produce better results. The complex J0034 field indicates the need for advanced 3GC technique for the next generation of low frequency and wide bandwidth telescopes. Finally, we aim to carry out statistical studies on the full sample once the narrowband and wideband analysis is completed. 

\vspace{6pt} 


\authorcontributions{Conceptualization, K.M. and K.K.; methodology, K.M., K.K. and S.P.S.; software, C.H.I.-C. and M.H.; formal analysis, S.P.S.; writing---original draft preparation, S.P.S.; writing---review and editing, M.H., N.G., K.K. and K.M.; supervision, K.M., K.K. and M.H. All authors have read and agreed to the published version of the manuscript.}

\funding{S.P.S. acknowledges funding from the South African Radio Astronomy Observatory and the National Research Foundation (NRF Grant Number: 95533).}

\institutionalreview{Not applicable}

\informedconsent{Not applicable}

\dataavailability{The uGMRT data reported in this paper are available through the GMRT online archive.
 \url{https://naps.ncra.tifr.res.in/goa}. The data were last accessed November 2020. Besides the raw UVFITS data, SPAM pipeline processed GSB images are also available in the GMRT archive. The observation IDs for the gsb and gwb datasets are 8651 and 9434, respectively.} 

\acknowledgments{S.P.S. acknowledges funding support from NRF/South African Radio Astronomy Observatory. We thank the staff of the GMRT who made these observations possible. GMRT is run by the National Centre for Radio Astrophysics of the Tata Institute of Fundamental Research. CHIC acknowledges the support of the Department of Atomic Energy, Government of India, under the project  12-R\&D-TFR-5.02-0700.}

\conflictsofinterest{The authors declare no conflict of interest.}

\end{paracol}
\printendnotes[custom]
\reftitle{References}


\end{document}